Steven Diamond

Stephen Boyd

David Greenberg

Mykel Kochenderfer

Andrew Ang

# Optimal Claiming of Social Security Benefits

**This Version: May 19, 2021**

Using a lifecycle framework with Epstein-Zin (1989) utility and a mixed-integer optimization approach, we compute the optimal age to claim Social Security benefits. Taking advantage of homogeneity, a sufficient statistic is the ratio of wealth to the primary insurance amount (PIA). If the investor's wealth to PIA ratio exceeds a certain threshold, individuals should defer Social Security for at least a year. The optimal threshold depends on mortality assumptions and an individual's utility preferences, but is less sensitive to capital market assumptions. The threshold wealth to PIA ratio increases from 5.5 for men and 5.2 for women at age 62 to 11.1 for men and 10.4 for women at age 69. Below the threshold wealth to PIA ratio, individuals claim Social Security to raise consumption. Above this level, investors can afford to fund consumption out of wealth for at least one year, and then claim a higher benefit.

Acknowledgments: The views expressed here are those of the authors alone and not of BlackRock, Inc. We thank Emmanuel Candes, Partha Mamidipudi, Vincent Cocula, and Fiona Sloof for helpful comments. The corresponding author is Steven Diamond, who can be reached by email at steven.diamond@blackrock.com.

Social Security provides inflation-indexed life annuities starting from ages 62 to 70, with the amounts received depending on work history, family status, and age. Social Security benefits represent the majority of income received by retired Americans. Dushi, Iams, and Trenkamp (2017) calculate that at least half of Americans over 65 receive at least 50% of their income from Social Security and for about one in four senior households, Social Security provides at least 90% of income. Not only is the income from Social Security essential for most Americans, it also constitutes around 40% of total household savings for working Americans (see Gustman, Steinmeier, and Tabatabai, 2010; Hubener, Maurer, and Mitchell, 2016).

The amounts in the Social Security benefits stream depend on the age of the recipient at the date of claim. For persons born in 1960 and after, the full retirement age (FRA) is 67 and at that age, recipients electing Social Security receive 100% of their primary insurance amount (PIA). The PIA is a proportion of average indexed monthly earnings, up to a maximum, with the proportions depending on the year in which a worker attains age 62 (or becomes disabled before 62 or dies before age 62 if those benefits are claimed by a spouse). But, a person born in 1960 or later and retiring at age 62 receives only 70% of the PIA. On the other hand, delaying retirement to age 70 increases the benefits to 124% of the PIA.[1] The benefits depend on both the birth year of an individual and the time they elect to take Social Security. Not surprisingly given the importance of Social Security for income in retirement, Shoven and Slavov (2014a) claim that exercising the option to start receiving Social Security benefits is one of the most important, and complex, decisions facing Americans.[2]

In this paper, we compute the optimal age to claim Social Security benefits. We characterize the optimal Social Security decision as income and wealth levels change, relative to the baseline case without Social Security. That is, the baseline case is a lifecycle model following Samuelson (1969), Merton (1969), and Bodie, Merton, and Samuelson (1992) that sets asset allocation over equities and bonds and solves for optimal consumption as a function of age and income.[3] We assume labor income is zero in retirement (or decumulation), and solve the model following Moehle and Boyd (2021) who develop a deterministic convex optimization problem which matches the base lifecycle problem and includes stochastic income, asset returns, and mortality. With suitable parameters, the lifecycle model can match any glide path of risky and risk-free assets—including actual investible target date funds. As a baseline case to determine the effects of the Social Security election, we calibrate the baseline to the glide path in the BlackRock LifePath funds.[4]

---

[1] See https://www.ssa.gov/oact/ProgData/ar_drc.html. The increase in benefits is designed to be actuarially fair (see Sass, Sun, and Webb, 2007).
[2] Purcell (2020) shows that over 2015 to 2018, 22% of men aged 62 claimed Social Security and 25% of women. There has been a pronounced downward trend in those aged 62 claiming Social Security even as the total number of eligible people has increased. Before 2000, the proportion of men and women claiming Social Security at age 62 were 44% and 49%, respectively. This is attributable, to among other factors, changes to Social Security in 1983 which reduced benefits payable before reaching the FRA and also increased the FRA for younger persons.
[3] See Ang (2014) for a literature summary of lifecycle models.

We characterize the solution by computing a critical wealth to PIA ratio which varies with age. At age 62, a woman should claim Social Security that year if her wealth to PIA ratio is below 5.2. In this case, she claims Social Security and raises her consumption immediately. If her ratio of wealth to PIA is above 5.2, she should defer claiming Social Security. In terms of utility, she funds consumption from her savings for one or more years to enjoy a higher Social Security benefit later. At age 62, a man's critical wealth to PIA ratio is 5.5. The threshold for men is higher due to increased mortality. The threshold wealth to PIA ratios increase to 11.1 for men and 10.4 for women at age 69.

We examine how these results change as we vary investment returns, risk aversion, and the intertemporal elasticity of substitution. Of particular interest is the sensitivity of our results to changes in expected returns—especially as low real rates have become pervasive and, over the last two decades, nominal interest rates have fallen more than the rate of inflation (see, among others, Andrade et al., 2019; Eberly, Stock, and Wright, 2019). We find that only for unrealistically high expected returns would individuals wish to claim Social Security immediately. In this case, the higher benefits are worth less in present value terms and the immediate Social Security benefit may be invested in an investment portfolio with a very high return. Consistent with Horneff, Maurer, and Mitchell (2018), for realistic expected returns, we find that individuals wish to defer Social Security as long as their wealth to PIA ratio is above a critical threshold.

Our study belongs to a literature examining the optimal claiming age for Social Security. Our paper falls into the positive economics side of this literature, providing a framework to optimize the Social Security claim, like Coile et al. (2002), Munnell and Soto (2005), and Shoven and Slavov (2014a).[5] Most of this literature advocates delaying Social Security (until age 70) partly because the monthly benefit becomes permanently higher.[6] A notable exception is Manakyan, Ervin, and Claggett (2014) who advocate taking Social Security early so that the proceeds can earn higher returns in individuals' investment portfolios. We find little empirical evidence for this case. Our main contribution to this literature is to characterize the decision to claim Social Security based on individuals' wealth to PIA ratio, and show how the threshold wealth to PIA ratio varies with age, investment returns, and utility preferences.

---

[4] The first target date funds were launched in November 1993 by Wells Fargo Investment Advisors, which is now part of BlackRock.
[5] Normative empirical studies examining how people claim Social Security benefits include Coile et al. (2002), Hurd, Smith, and Zissimopoulos (2004), Sass, Sun, and Webb (2007), and Benitez-Silva and Heiland (2008).
[6] Other authors examining the optimal claiming age for Social Security benefits include Sun and Webb (2009), Knoll and Olsen (2014), Meyer and Reichenstein (2010, 2012a and b), Munnell, Golub-Sass, and Karamcheva (2013), Laster and Suri (2014), Alleva (2015), and Reichelstein and Meyer (2016).

The rest of the paper is organized as follows. We lay out the model in Section 1, which describes the lifecycle model problem formulation and how we solve the problem to optimally claim Social Security. We report the results of the model in Section 2, including various sensitivities of the optimal Social Security age election to asset returns and utility preferences. Section 3 concludes.

# 1. Model

Section 1.1 outlines a standard lifecycle model, which is the setting in which an agent optimally chooses to claim Social Security benefits between ages 62 and 70. Section 1.2 describes the optimization technique.

### *1.1. Lifecycle specification*

Our setting is a standard lifecyle model where, as is appropriate for the retirement setting, there are no income sources other than investment returns and Social Security. For simplicity, we model a single risky asset and an unmarried individual born after 1960; for individuals born before 1960, there are different schedules for the amount of Social Security as a function of PIA, but the same methodology can be used. The parameters of the lifecycle model are the rate of mortality, expected inflation-adjusted portfolio return and volatility, utility preferences, starting age, starting wealth, and PIA. We take the rate of mortality, portfolio return and volatility, and risk aversion from the lifecycle model underlying the BlackRock LifePath funds (see O'Hara and Daverman, 2015).[7]

The lifecycle model is as follows. The state of the agent in the lifecycle model at age $t$ is the total wealth $w_t$, PIA, and whether Social Security has been claimed (and if so, at what age). The agent receives Social Security income $s_t$ based on the PIA and claiming age. This means that $s_t = 0$ if Social Security has not been claimed. The agent consumes $c_t$.

The agent receives investment return $r_t$ on wealth $w_t$. The return $r_t$ is a normal random variable with mean $\mu$ and variance $\sigma^2$:

$$r_t \sim \mathcal{N}(\mu, \sigma^2). \tag{1}$$

The wealth dynamics are given by:

---

[7] We use the RP-2014 table from the Society of Actuaries for mortality probabilities. Our inflation-adjusted capital market assumptions for the 60/40 equity/bond portfolio are 2.89% return and 9.65% volatility. Our EZ parameters are $\delta = 1, \rho = -2.5$, and $\gamma = -2.6$. We explore the effects of varying the capital market assumptions and EZ parameters in Section 2.

$$w_{t+1} = (1 + r_t)w_t + s_t - c_t. \tag{2}$$

In words, wealth increases by the return on wealth at the beginning of the period plus the Social Security benefit less consumption. Our decision variables are both the Social Security income, $s_t$—where we decide on the optimal age to claim Social Security—and the individual's consumption, $c_t$.

We model utility preferences in terms of Epstein-Zin (1989) (EZ) utility:

$$U_t(c_t, w_t) = \left\{ c_t^\rho + \delta E_t \left[ p_{t+1|t} \, U_{t+1}(c_{t+1}, w_{t+1})^\gamma \right]^{\frac{\rho}{\gamma}} \right\}^{\frac{1}{\rho}}, \tag{3}$$

where $\delta$ determines the marginal rate of time preference, $\rho$ determines the intertemporal elasticity of substitution, and $\gamma$ determines the investor's risk aversion. We take into account mortality with the probability $p_{t+1|t}$, which is the probability the agent continues to age $t+1$ conditional on attaining age $t$.

The agent chooses consumption and whether or not to elect Social Security each year. We make two important simplifying assumptions. First, we assume that the decision to claim Social Security is irrevocable—that is, once elected, the investor receives the stream of Social Security income. In reality, the investor has an option to claim Social Security and then file to suspend the benefits ("claim and suspend").[8] During the suspension, the individual accrues retirement credits which increase the PIA. This could be optimal if an individual, for example, claims Social Security early to maintain consumption and then comes into an unexpected windfall (say through an inheritance). We rule out this case because we assume there is no income other than from investment returns or Social Security.

Second, in our baseline case we focus on the claiming decision of Social Security and abstract away from any portfolio allocation: any wealth not consumed is automatically invested in a 60/40 stock-bond portfolio that follows returns $r_t$. EZ utility, like that of widely used Constant Relative Risk Aversion (CRRA, which is a special case of EZ), produces optimal portfolio weights that are constant under iid returns (see, for example, Merton, 1969; Samuelson, 1969). In a second set of exercises, we change the expected return on the individual's portfolio and examine the effect on the optimal age to claim Social Security.

---

[8] See ssa.gov/benefits/retirement/planner/suspend.html. Some common strategies concern married couples. Primary earners can delay Social Security while the spouse makes an early claim. The primary earner's PIA continues to grow and allows the spouse to claim spousal benefits when the primary has reached FRA. If the primary earner dies, the widow is entitled to the primary earner's PIA plus delayed retirement credits the primary earner was entitled to at time of death. For other strategies, see Shuart, Weaver, and Whitman (2010), Munnell, Golub-Sass, and Karamcheva (2013), and Shoven and Slavov (2014a and b). We leave spousal benefits to future work and limit ourselves in this paper to an individual problem.

### *1.2 Optimization and homogeneity*

We optimize the lifecycle model using a mixed-integer extension of Moehle and Boyd (2021). Traditionally, dynamic programming approaches have been used to solve the lifecycle model (equations (1)-(3)) like Viceira (2001), Campbell and Viceira (2002), and Cocco, Gomes, and Maenhout (2005). Moehle and Boyd (2021) develop a convex optimization solution that is equivalent to the dynamic programming problem but is much more computationally efficient. We model the decision to claim Social Security at each age by representing it as nine binary variables, one for each age between 62 up to and including 70. (The last year to claim Social Security is 70, at which point there is no benefit in delaying further.) The relevant outputs of the optimization are what fraction of wealth to consume at the starting age and when to claim Social Security.

Following Moehle and Boyd (2021), we optimize the lifecycle model by solving the optimization problem:

$$\text{maximize } \sum_{t=t_0}^{T} \delta^{t-t_0} \left( \prod_{i=t_0}^{t-1} p_{i+1|i} \right) (c_t)^\rho / \rho \tag{4}$$
$$\text{subject to } w_{t+1} \leq (1+\mu)w_t - c_t + s_t + \frac{(\gamma-1)}{2} \frac{w_t^2 \sigma^2}{w_t + \sum_{i=t}^{T} s_i}, \quad t = t_0, \ldots, T$$
$$\theta_t \in \{0,1\}, \quad t = 62, \ldots, 70$$
$$\sum_{t=62}^{70} \theta_t = 1$$
$$s = PIA \sum_{t=62}^{70} \theta_t v^{(t)}$$
$$c_t \geq 0, \quad t = t_0, \ldots, T$$
$$w_t \geq 0, \quad t = t_0, \ldots, T+1$$

where $t_0$ denotes the starting age, $T$ the maximum terminal age, and $v^{(t)}$ the cash flow vector associated with claiming Social Security at age $t$. The optimization variables are $c_{t_0}, \ldots, c_T$, $w_{t_0+1}, \ldots, w_T$, $s_{t_0}, \ldots, s_T$, and $\theta_{62}, \ldots, \theta_{70}$. The problem inputs are $\rho, \gamma, \delta, \mu, \sigma^2, PIA, w_{t_0}$ and the probabilities $p_{t+1|t}$. The solution to the optimal age to claim Social Security is the age $t$ where $\theta_t = 1$.

Problem (4) is non-convex, but can be easily solved by mixed-integer program solvers. The problem has a particularly simple solution because the binary variables $\theta_{62}, \ldots, \theta_{70}$ can only take on nine different values, representing the possible Social Security claiming ages. When the values of $\theta_{62}, \ldots, \theta_{70}$ are fixed, problem (4) becomes convex and can be solved efficiently using standard methods (see Boyd and Vandenberghe 2004). We solve problem (4) by trying all nine possible values of of $\theta_{62}, \ldots, \theta_{70}$ (equivalently, every possible claiming age) and seeing which results in the highest objective value.

The lifecycle model is homogeneous with respect to the ratio of initial wealth to PIA. In other words, the optimization outputs do not depend on initial wealth and PIA separately, but on their ratio. It follows that the decision criteria for whether to claim Social Security at any given age can be summarized simply: if the ratio of wealth to PIA is below a certain threshold, it is optimal to claim Social Security at that age. If the ratio is above the threshold, it is optimal to defer until a later age. We compute the threshold by optimizing the lifecycle model and finding the cutoff point where the investor is indifferent between claiming Social Security immediately or deferring.

## 2. Optimal Election of Social Security Benefits

We present the empirical analysis in this section. We start in Section 2.1 with details on how the Social Security benefit is calculated. In Section 2.2, we give the intuition behind the solution of determining the optimal age to claim Social Security. The intuition frames the main empirical results in Section 2.3. Sections 2.4 to 2.6 present comparative static exercises, showing how the optimal claiming age changes as investment returns, risk aversion, and the intertemporal elasticity of substitution change, respectively.

### *2.1. Mechanics of Social Security Benefit calculation*

The basic Social Security benefit is called the primary insurance amount (PIA). The PIA is computed from the employee's average indexed monthly earnings (AIME). The AIME is computed from the average wage-index adjusted monthly earnings of the employee based on their highest 35 earning years adjusted for wage-inflation.[9] Based on 2020 rules, the PIA is defined as the level to replace 90% of AIME up to $906 per month, 32% of AIME up to $5,785 per month, and 15% of AIME for AIME in excess of $5,785 per month. In this way, the PIA replaces almost all income for workers in low income brackets but replaces a lower fraction of working income at higher wage levels.

The PIA is calculated at age 62 and reflects the monthly retirement benefit assuming retirement at the full retirement age (FRA) and is adjusted thereafter for the cost of living (cost of living adjustments, or COLA). We ignore in this paper the possibility of the PIA changing after age 62 due to new labor income. The FRA for anyone born after 1960 is 67 years and decreases by two months per year for birth years before 1960. Individuals can elect to begin Social Security payments as early as age 62 or as late as age 70, but the Social Security benefit as a fraction of PIA is reduced (increased) on a sliding scale for benefits starting before (after) the FRA. Taking Social Security benefits starting at 62 translates to monthly payments that are 70% of PIA given an FRA of 67; taking benefits at age 70 translates to monthly payments that are 124% of PIA given an FRA of 67. The first row of Exhibit 1 shows the full schedule of Social Security benefits as a percentage of PIA by age for an individual born after 1960. The second row shows the percent increase in benefit from delaying election from the previous year. Social Security benefits are capped; the maximum monthly Social Security benefit assuming retirement at FRA in 2021 is $3,148. When claiming benefits at age 70, the maximum benefit increases to $3,985.

---

[9] The numeraire in computing the wage index for each year is the average national wage at age 60.

If the objective in choosing the age to begin receiving Social Security is to maximize the total expected benefit payout, the calculation is purely an actuarial one. Delaying the payout increases the monthly benefit but decreases the total number of monthly payments based on expected mortality. Exhibit 2 shows the expected cumulative Social Security benefit for each claiming age between 62 and 70 relative to the expected cumulative benefit for a male claiming at age 62. The cumulative payout increases with the claiming age as the benefit from a higher payout per month outweighs the shorter stream of payments. As expected from their lower mortalities, women have a higher total sum of benefits and the benefit gap widens with the age; a woman who claims Social Security benefits at age 70 should expect 32% higher cumulative benefits than a man who claims benefits at age 62.

## 2.2. Intuition

Optimizing the year in which to begin Social Security benefits can be reduced to the problem of evaluating at each year after age 62 whether one can afford to defer benefits another year and consume out of current savings. Unless the forecasted Sharpe ratio of the investor's portfolio is extremely high, our model shows that the optimal decision is to defer Social Security each year and draw down on the portfolio to meet current consumption. The optimal policy reduces to a simple rule of comparing the ratio of current wealth to the the PIA benefit. Above this threshold ratio, the investor should defer Social Security for at least another year, and below this ratio the investor should claim Social Security immediately.

Consider an individual aged 62 with PIA equal to $1,000. The immediate decision is whether to take benefits now at 62 or defer. If that individual elects benefits at 62, monthly benefits are 70%×PIA = $700 per month. If the individual defers and elects benefits at 63, monthly benefits will be 75%×PIA = $750 per month. By deferring a year, an individual gives up $8,400 in benefits at age 62 in exchange for a guaranteed COLA-annuity of $50 per month starting at age 63. That is, deferring for one year is economically equivalent to giving up an annuity starting today of $700 per month to purchase an annuity starting next year paying $750 per month. Taking the difference between these two annuities, or foregoing $700 per month this year, generates an annuity benefit of $50 per month plus COLA adjustments starting next year. The actuarial present value of an inflation-indexed life annuity of $1 per year at age 62 is approximately $25 given current interest and inflation rates.[10] So, the fair value of the $50 per month annuity is $25 × $5,012 × 12 = $15,000, which is nearly double the implicit cost of $8,400 of foregoing benefits for a year.[11] When the Social Security benefits schedule in Exhibit 1 was first designed, the increase in payout by deferring benefits was calibrated to be actuarially fair; however the decline in interest rates combined with decreasing mortality have strongly tilted the calculation in favor of deferring benefits.

Despite the higher benefits from delaying, there are two scenarios where it would be optimal to claim Social Security benefits immediately. First, if returns to the risky portfolio are high enough, it may be optimal to elect $700 per month in benefits at age 62, invest the $8,400 in benefits in the risky portfolio, and withdraw a COLA-adjusted amount of $50 per month starting at age 63. However, the portfolio would have to generate very high *real* returns with low volatility to avoid a high probability of wealth going to zero and the individual outliving her assets. We explore scenarios with different portfolio returns in section 2.3, but our default capital market assumptions are inconsistent with such a high return, low volatility portfolio.[12]

The second reason to opt for benefits at age 62 is that initial wealth is so low that the individual needs retirement income immediately to meet present consumption needs. No matter how high future benefits are, claiming immediate benefits is optimal if initial wealth is low enough. In our example, if initial wealth is $9,000, the individual can completely draw down savings and consume $750 per month at age 62, defer Social Security for one year, and then live off the COLA-adjusted $750 per month Social Security benefit starting at age 63. Generalizing for any level of PIA, if $W/PIA \geq 9$, the individual has enough wealth to have a constant consumption profile over time and defer Social Security for at least one year at age 62. If $W/PIA < 9$, the individual will have to consume less at age 62 than subsequent years if she defers Social Security. The $W/PIA$ threshold at which the individual can defer Social Security and still have constant consumption is plotted in Exhibit 2 as a dotted green line. The $W/PIA$ threshold increases with age since benefits as a fraction of PIA increase with age. This $W/PIA$ threshold is the optimal policy rule only for those individuals who demand smooth consumption over time. Individuals that can tolerate lower consumption now in favor of higher consumption later will have a $W/PIA$ threshold strictly below this line.

---

[10] Refer to the Society of Actuaries annuity factor calculator at https://afc.soa.org/
[11] In addition, there is the option to defer yet again at age 63, increasing the value of the annuity further.

[12] Alternatively, the individual could invest the $8,400 in an inflation indexed single premium immediate annuity (SPIA). However, they would only be able to purchase a $8,400/$300 = $28 per month annuity.

### *2.3. Optimal Ages to Claim Social Security*

The solid red line in Exhibit 3 shows the $W/PIA$ threshold after optimizing the consumer's EZ utility to balance the demand for smoothed intertemporal consumption and higher lifetime benefits. Notice that the solid red line is below the dotted green line: that is, the individual is willing to take significant lower consumption (roughly 1/3 lower) for one year in order to get higher lifetime Social Security benefits. Increasing the intertemporal elasticity of substitution shifts down the $W/PIA$ threshold in Exhibit 3.

In the full EZ lifecycle problem, the critical $W/PIA$ red line is not smooth; it ratchets up discontinuously as age increases. This ratcheting is caused by the variability in the percentage increase in lifetime benefits by age. As shown in Exhibit 1, the percentage increase in lifetime benefits from deferring Social Security ranges from 6.7% to 8.3%. When the benefit increase is large, e.g. the benefit increases by 8.3% between ages 64 and 65, the threshold $W/PIA$ is lower, reflecting the optimal decision to take even lower consumption for one year to lock in the greater increase in benefits (i.e. the higher implicit annuity value).

We can illustrate the effect of the PIA increases in age (see Exhibit 1) inducing ratcheting in the the optimal $W/PIA$ ratios by computing the thresholds if we counterfactually assume the PIA increases by a constant 7.5% for every age. We plot this as a black dotted line in Exhibit 3, which is smooth and has a similar slope to the green dotted line. The ratio between the threshold $W/PIA$ of the black line and the $W/PIA$ constant consumption line increases slightly with age, which reflects the decreased actuarial fair value of the embedded annuity as mortality increases with age. At ages 62 and 69, the ratio between the optimal and constant consumption $W/PIA$ thresholds are 57% and 66%, respectively. In other words, the agent requires a higher level of immediate consumption as a fraction of expected future consumption at older ages to offset the lower benefit from the embedded annuity.

Exhibit 4 shows the $W/PIA$ threshold is lower for women than for men since they have lower rates of mortality and the value of the embedded annuity is correspondingly higher. The $W/PIA$ thresholds for women and men are 5.2 and 5.5 respectively at age 62. By age 69, the $W/PIA$ thresholds for women and men respectively increase to 10.4 and 11.1.

### *2.3. Effect of Investment Returns*

In the examples presented so far, deferring Social Security to age 70 was optimal provided that the agent has sufficiently high wealth. At low levels of $W/PIA$, the decision to exercise early depends on the trade-off between the value of the extra benefit from delaying an extra year compared to the cost of consuming for one year at a level lower than future years.

The increased benefit from delaying Social Security, however, is partially offset by the foregone investment return of the benefit if it is claimed early. The larger the expected return on the investment portfolio, the stronger the incentive to claim early. Exhibit 5 plots the threshold $W/PIA$ as a function of different portfolio returns. In this exercise, we fix the portfolio volatility at 10% and alter only the investment return. Exhibit 5 uses a logarithmic scale as there are large changes in the ratios for different expected returns. Consider all expected returns to be real, i.e. adjusted for inflation.

As expected, at higher levels of expected return, the agent is willing to exercise early at higher levels of the $W/PIA$ threshold. It is interesting, however, that the impact on the $W/PIA$ threshold is extremely modest for levels of inflation-adjusted return below 7%. At levels of return below 7%, and at wealth levels slightly above the $W/PIA$ threshold, the agent fully consumes from current wealth and defers Social Security; the return of the portfolio is irrelevant because the wealth endowment is fully consumed in that period. At levels slightly below the $W/PIA$ critical value, the agent claims benefits immediately, and an increase in the return on the wealth portfolio increases future consumption, making claiming benefits early slightly more attractive than before. This pushes up the $W/PIA$ line modestly.

At levels of return at 7% and above, however, the agent may achieve a higher level of utility (i.e. higher risk-adjusted consumption) by claiming benefits early and investing them in the risky asset. The return stream from the risky asset has become more valuable than the increased benefit stream by delaying Social Security, and agents should claim early at much higher levels of $W/PIA$ than previously. If the level of the real return is high enough, almost all individuals should elect at age 62 because the benefit from investing early in the risky asset more than outweighs the value of the increase in benefits from delaying election.

Estimates of the real rate have decreased by a few percentage points since the early 2000s (see Chung et al., 2019 for seven estimates of the real rate), and 10-year TIPS are trading close to zero or negative. Furthermore, many forward-looking estimates of expected returns over the next decade are much lower than over previous decades (see, for example, Ren et al., 2021). Thus, it is most likely that individuals should defer Social Security if their wealth is sufficiently high. It is also notable that the optimal claiming ages are not sensitive to changes in expected returns, except for extremely high expected returns.

### 2.4. Effect of Risk Aversion

Risk aversion has minimal effect for inflation-adjusted returns below 7%. At levels of return below 7%, and at wealth levels slightly above the $W/PIA$ threshold, the agent fully consumes from current wealth and defers Social Security; the risk of the portfolio does not matter since the agent fully consumes their wealth endowment in that period. Only for returns greater than 7% does higher risk aversion become important. In Exhibit 6, we graph the threshold $W/PIA$ ratio for investors with different risk aversions, assuming a portfolio with 7% return and 10% volatility. As risk aversion increases, investors are less inclined to claim early and invest their benefits because the greater expected return is offset by the precautionary risk. The higher risk aversion therefore lowers the $W/PIA$ threshold at which claiming early becomes optimal.

### 2.5. Effect of Intertemporal Elasticity of Substitution

A feature of EZ utility is that, unlike the classical CRRA utility, the effect of risk aversion is separate from the intertemporal elasticity of substitution (IES). This may matter for older individuals who may prefer to endogenenously increase mortality risk for small values of IES when consumption is small (see, for example, Marshall, 1984; Rosen, 1988). In Exhibit 7, we graph the threshold $W/PIA$ ratio for investors with different IES. As IES increases, investors are more willing to have lower levels of consumption in one year if it is more than offset by higher consumption in the other years. At very low levels of IES, the optimal $W/PIA$ ratio converges to the constant consumption line, as investors demand smoother and smoother consumption profiles.

# 3. Conclusion

We build a model to compute the optimal age to claim Social Security benefits. We find, that except in the unlikely case of extremely high expected returns, an individual should defer Social Security to age 70 if their wealth is sufficiently high. In our baseline calibrations, a man turning 62 should defer Social Security for at least one year if his wealth to primary insurance amount (PIA) ratio is at least 5.5. A woman turning 62 should defer if her wealth to PIA ratio is over 5.2. These ratios of wealth to PIA increase to 11.1 and 10.4 at age 69 for men and women, respectively. Individuals should claim Social Security if their wealth falls below the critical wealth to PIA threshold, and defer if their wealth lies above this threshold. Intuitively, for wealth below this threshold, individuals claim Social Security to ensure smooth consumption. For high enough wealth levels, the individual can afford to defer Social Security, consume out of wealth for one year, and then claim higher Social Security benefits one year later.

There are a number of extensions of our results. Although we worked with Epstein-Zin (1989) utility, it is likely these results will apply to other utility functions because the main driver behind the optimal deferral of Social Security is the trade-off between higher expected lifetime benefits and smoother consumption. Behavioral utility functions, however, may change this trade-off. Our analysis applies only to an individual and the decision to claim joint Social Security benefits, especially for survivor benefits, is much more complicated. Finally, we have assumed a constant expected return on investments, which is the same as assuming the stock-bond mix is held constant. Extensions of our analysis to optimal asset allocation in the presence of Social Security, following Hubbard (1985), Reichenstein (2000), and Fraser, Jennings, and King (2000), we leave to future research. In contrast with existing work, our methodology allows the asset allocation to be determined when the Social Security claiming age is endogenous.

## References

Alleva, B. J., 2015. Minimizing the Risk of Opportunity Loss in the Social Security Claiming Decision, Journal of Retirement 3, 67-86.

Andrade, P., J. Gali, H. Le Bihan, and J. Matheron, 2019. The Optimal Inflation Target and the Natural Rate of Interest, Brookings Papers on Economic Activity, vol. 2019(2), 173-255.

Ang, A. 2014. Asset Management: A Systematic Approach to Factor Based Investing. New York, NY: Oxford University Press.

Benitez-Silva, H., and F. Heiland, 2008. Early Claiming of Social Security Benefits and Labor Supply Behavior of Older Americans, Applied Economics 40, 2969-2985.

Bodie, Z., R. C. Merton, and W. F. Samuelson, 1992, Labor Supply Flexibility and Portfolio Choice in a Life-Cycle Model, Journal of Economic Dynamics and Control 16, 427-449.

Boyd, S., and L. Vandenberghe, 2004. Convex Optimization. New York, NY: Oxford University Press.

Campbell, J. Y., and L. M. Viceira, 2002. Strategic Asset Allocation: Portfolio Choice for Long-Term Investors. New York, NY: Cambridge University Press.

Chung, H., E. Gagnon, T. Nakata, M. Paustian, B. Schlusche, J. Trevino, D. Vilan, and W. Zheng, 2019. Monetary Policy Options at the Effective Lower Bound: Assessing
the Federal Reserve's Current Policy Toolkit, Finance and Economics Discussion Series 2019-03.

Cocco, J. F., F. J. Gomes, and P. J. Maenhout, 2005. Consumption and Portfolio Choice over the Life-Cycle, Review of Financial Studies 18, 491-533.

Coile, C., P. A. Diamond, J. Gruber, and A. Jouston. 2002. Delays in Claiming Social Security Benefits, Journal of Public Economics 84, 357-85.

Dushi, I., H. M. Iams, and B. Trenkamp, 2017. The Importance of Social Security Benefits to the Income of the Aged Population, Social Security Bulletin 77, 1-12.

Eberly, J. C., J. H. Stock, and J. H. Wright, 2019. The Federal Reserve's Current Framework for Monetary Policy: A Review and Assessment, Working Paper.

Epstein, L., and S. Zin, 1989. Substitution, Risk Aversion, and the Temporal Behavior of Consumption and Asset Returns: A Theoretical Framework, Econometrica 57, 937-969.

Fraser, S. P., W. W. Jennings, and D. R. King, 2000. Strategic Asset Allocation for Individual Investors: The Impact of the Present Value of Social Security Benefits 9, 295-326.

Gustman, A. L., T. L. Steinmeier, and N. Tabatabai, 2010. What the Stock Market Decline Means for the Financial Security and Retirement Choices of the Near-Retirement Population, Journal of Economic Perspectives 24, 161–82.

Hubbard, R. G., 1985. Personal Taxation, Pension Wealth, and Portfolio Composition, Review of Economics and Statistics 67, 53–60.

Hurd, M. D., J. P. Smith, and J. M. Zissimopoulos, 2004. The Effects of Subjective Survival on Retirement and Social Security Claiming, Journal of Applied Econometrics 19, 761-775.

Horneff, V., R. Maurer, and O. S. Mitchell, 2018. How Persistent Low Expected Returns Alter Optimal Life Cycle Saving, Investment, and Retirement Behavior, NBER Working Paper.

Hubener, A., R. Maurer, and O. S. Mitchell, 2016. How Family Status and Social Security Claiming Options Shape Optimal Life Cycle Portfolios, Review of Financial Studies 29, 937-978.

Knoll, M. A. Z., and A. Olsen, 2014. Incentivizing Delayed Claiming of Social Security Retirement Benefits Before Reaching the Full Retirement Age, Social Security Bulletin 14, 21-43.

Laster, D., and A. Suri, 2014. When Should You Claim Social Security? Journal of Retirement 2, 14-22.

Manakyan, H., D. Ervin, and E. T. Claggett, 2014. Take the Money: Should You Draw Social Security Retirement Benefits Early? Retirement Management Journal 4, 45-53.

Marshall, J. M., 1984. Gambles and the Shadow Price of Death, American Economic Review 74, 73–86.

Merton, R. C., 1969, Lifetime Portfolio Selection Under Uncertainty: The Continuous-Time Case, Review of Economics and Statistics 51, 247-257.

Meyer, W., and W. Reichenstein, 2010. Social Security: When Should You Start Benefits and How to Minimize Longevity Risk, Journal of Financial Planning 23, 52-63.

Meyer, W., and W. Reichenstein, 2012a. Social Security Claiming Strategies for Singles, Retirement Management Journal 2, 61-66.

Meyer, W., and W. Reichenstein, 2012b. How the Social Security Claiming Decision Affects Portfolio Longevity, Journal of Financial Planning 25, 53-60.

Moehle, N., and S. Boyd, 2021. A Certainty Equivalent Merton Problem, Working Paper, https://arxiv.org/abs/2101.10510

Munnell, A. H., A. Golub-Sass, and N. S. Karamcheva, 2013. Understanding Unusual Social Security Claiming Strategies, Journal of Financial Planning 26, 40-47.

Munnell, A. H., and M. Soto, 2005. Why Do Women Claim Social Security Benefits So Early? Issue in Brief 35, Center for Retirement Research at Boston College.

Munnell, A. H., N. S. Orlova, and A. Webb, 2012. How Important is Asset Allocation to Financial Security in Retirement? Working Paper 2012-13, Center for Retirement Research at Boston College.

O'Hara, M., and T. Daverman, 2015. Rexamining "To vs. Through": What New Research Tells Us about an Old Debate, Journal of Retirement 2, 30-37.

Purcell, P. J., 2020. Employment at Older Ages and Social Security Benefit Claiming, 1980-2018, Research and Statistics Note No. 2020-01, Social Security Administration.

Reichenstein, W., 2000. Calculating the Asset Mix, Journal of Wealth Management 3, 20–25.

Reichenstein, W., and W. Meyer, 2016. Social Security Claiming Strategies for Widows and Widowers, Journal of Retirement 3, 77-86.

Ren, H., S. Siwinski, C. Yu, and A. Ang, 2021. Public Pension Portfolios in a World of Low Rates and Low Risk Premiums. Working Paper, BlackRock.

Rosen, S., 1988. The Value of Changes in Life Expectancy, Journal of Risk and Uncertainty 1, 285–304.

Samuelson, P. A., 1969. Lifetime Portfolio Selection by Dynamic Stochastic Programming, Review of Economics and Statistics 51, 239-246.

Sass, S. A., W. Sun, and A. Webb, 2007. Why do Men Claim Social Security so Early: Ignorance or Caddishness? Working Paper 2007–17, Center for Retirement Research at Boston College.

Shoven, J. B., and S. N. Slavov, 2014a. Does it Pay to Delay Social Security?" Journal of Pension Economics and Finance 13, 121–44.

Shoven, J. B., and S. N. Slavov, 2014b. Recent Changes in the Gains from Delaying Social Security, Journal of Financial Planning 27, 32-41.

Shuart, A. N., D. A. Weaver, and K. Whitman, 2010. Widowed Before Retirement: Social Security Benefits Claiming Strategies, Journal of Financial Planning 23, 45-53.

Sun, W., and A. Webb, 2009. How Much Do Households Really Lose By Claiming Social Security at Age 62? Working Paper 2009-11, Center for Retirement Research at Boston College.

Viceira, L. M., 2001. Optimal Portfolio Choice for Long-Horizon Investors with Nontradable Labor Income, Journal of Finance 56, 433-470.

## Exhibit 1: Primary Insurance Amount (PIA)

| Age | 62 | 63 | 64 | 65 | 66 | 67 | 68 | 69 | 70 |
|---|---|---|---|---|---|---|---|---|---|
| **Benefit as %PIA** | 70.0% | 75.0% | 80.0% | 86.7% | 93.3% | 100.0% | 108.0% | 116.0% | 124.0% |
| **% increase from prior year** | - | 7.1% | 6.7% | 8.3% | 7.7% | 7.1% | 8.0% | 7.4% | 6.9% |

Note: The table lists PIAs for individuals born after 1960.

**Exhibit 2: Expected Social Security Benefits at Different Claiming years**

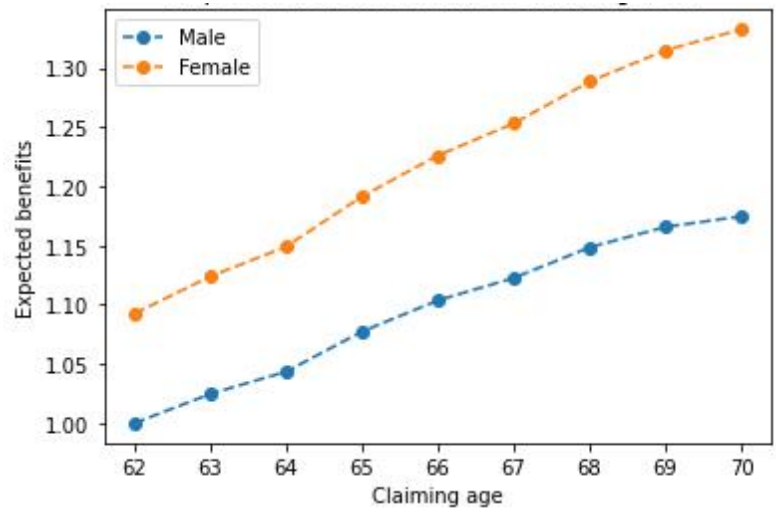

Note: 1.00 represents a male claiming Social Security benefits at age 62. All cumulative benefits are graphed relative to a male claiming at age 62.

**Exhibit 3: Wealth / PIA Threshold for Claiming Social Security**

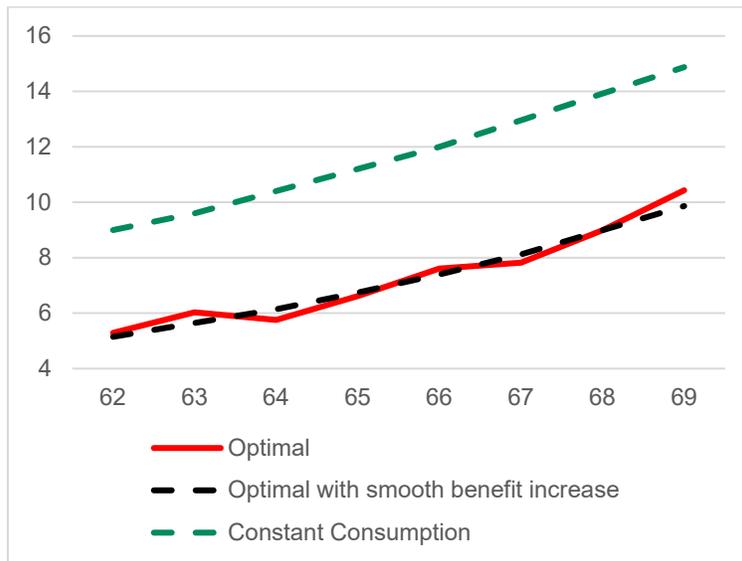

## Exhibit 4: Optimal Claiming Age and Gender

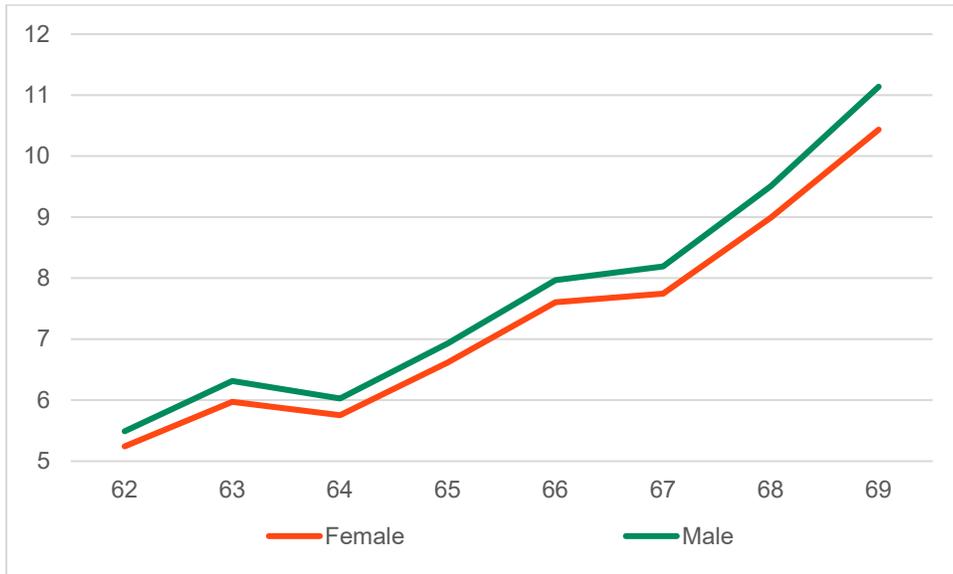

**Exhibit 5: Optimal Claiming Age and Investment Returns**

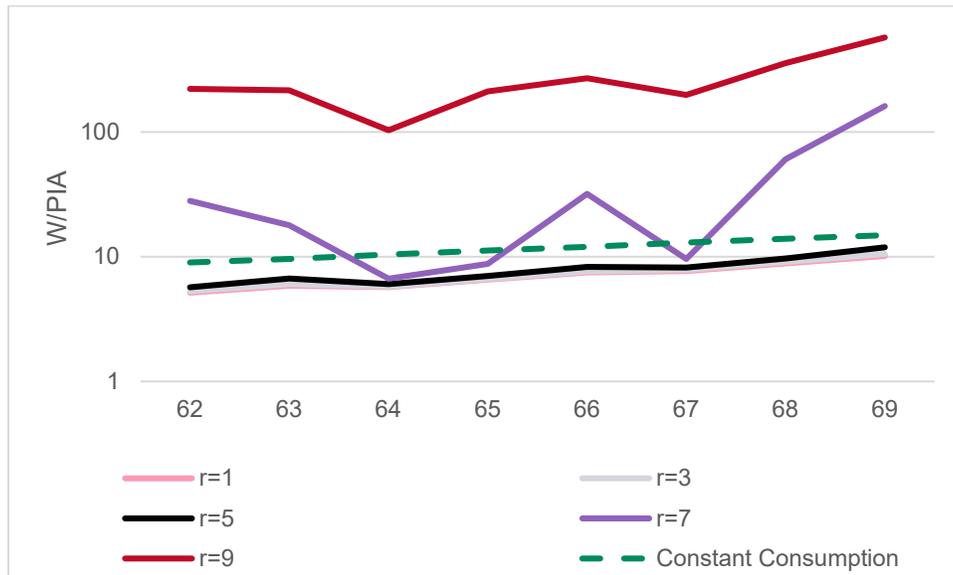

**Exhibit 6: Optimal Claiming Age and Risk Aversion**

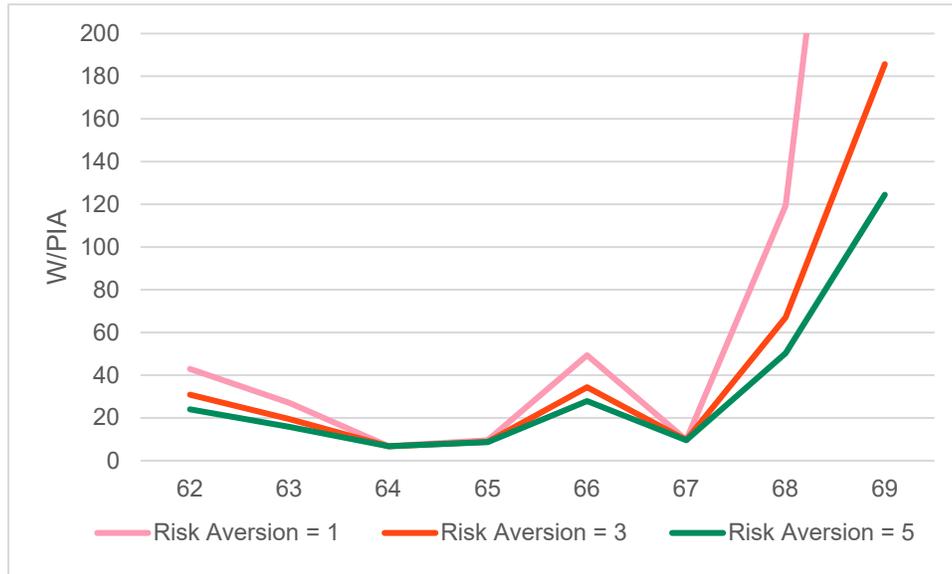

**Exhibit 7: Optimal Claiming Age and IES**

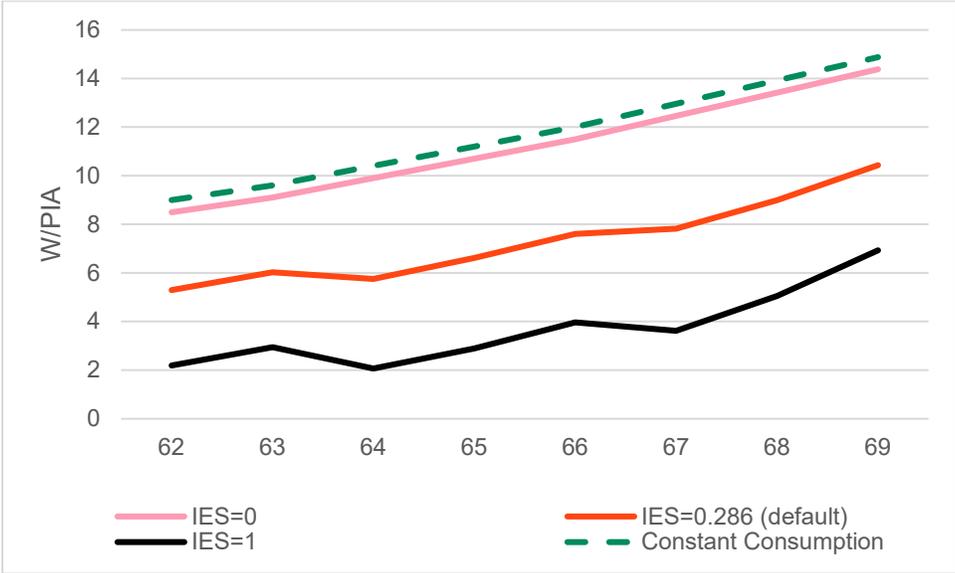

**Capital at risk.** All financial investments involve an element of risk. Therefore, the value of the investment and the income from it will vary and the initial investment amount cannot be guaranteed.

This material is provided for educational purposes only and is not intended to be relied upon as a forecast, research or investment advice, and is not a recommendation, offer or solicitation to buy or sell any securities or to adopt any investment strategy. The opinions expressed are subject to change. References to specific securities, asset classes and financial markets are for illustrative purposes only and are not intended to be and should not be interpreted as recommendations. Reliance upon information in this material is at the sole risk and discretion of the reader. The material was prepared without regard to specific objectives, financial situation or needs of any investor. This should not be construed as tax advice, please seek guidance from an accountant or other tax professional.

This material may contain "forward-looking" information that is not purely historical in nature. Such information may include, among other things, projections, forecasts, and estimates of yields or returns. No representation is made that any performance presented will be achieved by any BlackRock Funds, or that every assumption made in achieving, calculating or presenting either the forward-looking information or any historical performance information herein has been considered or stated in preparing this material. Any changes to assumptions that may have been made in preparing this material could have a material impact on the investment returns that are presented herein. Past performance is not a reliable indicator of current or future results and should not be the sole factor of consideration when selecting a product or strategy.

The information and opinions contained in this material are derived from proprietary and nonproprietary sources deemed by BlackRock to be reliable, are not necessarily all-inclusive and are not guaranteed as to accuracy.

This material is for Institutional use only – not for public distribution.

BlackRock® is a registered trademark of BlackRock, Inc., or its subsidiaries in the United States and elsewhere. All other trademarks are those of their respective owners.